\documentclass[a4paper,fleqn,usenatbib,useAMS]{mnras}

\usepackage{graphicx}	
\usepackage{amsmath}	
\usepackage{amssymb}	
\usepackage{multicol}        
\usepackage{bm}		
\usepackage{pdflscape}	

\newcommand{\kms}{\,km\,s$^{-1}$} 

\graphicspath{{Figures/}}
\newcommand{\totalCarina}{731}
\newcommand{\totalLR}{956}

\newcommand{\teff}{T$_{\rm eff}$}
\newcommand{\logg}{$\log g$}
\newcommand{\meta}{\hbox{[M/H]}}
\newcommand{\feh}{{\rm {[Fe/H]}}}

\def\kms{\,{\rm km~s^{-1}}}
\def\K{\,{\rm K}}
\def\cgs{\,{\rm cm~s^{-2}}}
\def\cgs{\,{\rm cgs}}
\def\kpc{\,{\rm kpc}}

\def\dex{\,{\rm dex}}
\def\nm{\,{\rm nm}}
\def\Gyr{\,{\rm Gyr}}

\def\ltsima{$\; \buildrel < \over \sim \;$}
\def\simlt{\lower.5ex\hbox{\ltsima}}
\def\gtsima{$\; \buildrel > \over \sim \;$}
\def\simgt{\lower.5ex\hbox{\gtsima}}

\usepackage[T1]{fontenc}
\usepackage{ae,aecompl}
\usepackage{times}

\title
  [Chemodynamic subpopulations of the Carina dwarf galaxy]{ Chemodynamic subpopulations of the Carina dwarf galaxy}
\author[G.~Kordopatis et al.]
  {G.~Kordopatis,$^1$\thanks{E-mail:
gkordopatis@aip.de}
    N.\,C.~Amorisco,$^{2,3}$
      N.\,W.~Evans,$^4$  G.~Gilmore,$^4$      S.\,E.~Koposov$^4$
   \\
  $^1$ Leibniz-Institut f\"ur  Astrophysik Potsdam (AIP), An der Sternwarte 16, 14482 Potsdam, Germany\\
  $^2$ Dark Cosmology Centre, Niels Bohr Institute, University of Copenhagen, Juliane Maries Vej 30, 2100 Copenhagen, Denmark\\
  $^3$ Max Planck Institute for Astrophysics, Karl-Schwarzschild-Strasse 1, D-85740 Garching, Germany\\
  $^4$ Institute of Astronomy, University of Cambridge, Madingley Road, Cambridge CB3 0HA, UK
}

\pagerange{\pageref{firstpage}--\pageref{lastpage}} \pubyear{2002}

\def\LaTeX{L\kern-.36em\raise.3ex\hbox{a}\kern-.15em
    T\kern-.1667em\lower.7ex\hbox{E}\kern-.125emX}

\begin{document}

\label{firstpage}

\maketitle

\begin{abstract}
We study the chemodynamical properties of the Carina dwarf
spheroidal by combining an intermediate spectroscopic resolution
dataset of more than 900 red giant and red clump stars, with
high-precision photometry to derive the atmospheric parameters,
metallicities and age estimates for our targets. Within the red giant
branch population, we find evidence for the presence of three distinct
stellar sub-populations with different metallicities, spatial
distributions, kinematics and ages. 
As in the Fornax and Sculptor dwarf spheroidals, the subpopulation with the lowest average metallicity is more extended and kinematically hotter than all other populations. However, we identify an inversion in the parallel ordering of metallicity, kinematics and characteristic length scale in the two {  most metal rich subpopulations, which therefore do not contribute to a global negative chemical gradient.} Contrary to common trends in the chemical properties with radius, the metal richest population is more extended and mildly kinematically hotter than the main component of intermediate metallicity.
More investigations are required to ascertain the nature of this inversion,
but we comment on the mechanisms that might have caused it.
\end{abstract}


\section{Introduction}

Since the late 1970s, it has been suggested that the Milky Way's halo
has been assembled from the merging and accretion of many smaller
systems \citep[e.g.][]{Searle78}. In this paradigm, it has long been
assumed that the Galaxy's satellites are remnants of this population of
building blocks.  However, first analyses of the chemical abundances
of the stellar populations in the surviving dwarf spheroidal galaxies
(dSphs) appeared inconsistent with this idea
{  \citep[e.g.,][]{Shetrone01,Venn04}}. In particular, the metal-poor tail of the
dSph metallicity distribution seemed significantly different from that
of the Galactic halo \citep{Helmi06}.  Since then, extremely
metal-poor stars have been discovered in Galactic satellite
galaxies such as Sculptor \citep[e.g.:][]{Frebel10}. Both the
classical and the ultra-faint dSphs are now known to have a wide
abundance dispersion and to host some of the most extreme metal-poor
stars known.  In fact, $\sim$ 30\% of the known stars with [Fe/H]
$\leq -3.5$ are found in dwarf galaxies
\citep[e.g.][]{Kirby10,Kirby12,Frebel10a}.

The connection between the surviving dwarfs and those that dissolved
to form the halo can partially be addressed by examining in detail the
stellar kinematics and chemical compositions of present-day dwarf
galaxies. Establishing the detailed chemical histories of these
systems can provide constraints on their dominant enrichment events
and timescales. In addition, modelling of the kinematics can yield
powerful information on the dark matter profile, especially in the
case of multiple populations \citep[e.g.:][]{Walker11, Amorisco12a}.

A multitude of independent studies have addressed the evolutionary
complexities of dwarf spheroidal galaxies in the Local Group.
Since \citet{Harbeck01}, stellar population gradients have been observed
in many satellites of both the Milky Way and Andromeda -- for example
Sculptor \citep{Tolstoy04}, Fornax \citep{Battaglia06}, Sextans \citep{Lee03},
Tucana \citep{Harbeck01}, Draco \citep{Faria07}, and And II \citep{Ho12}.
Whether such gradients are ubiquitous remains an open question, as
dwarfs like Leo I, And I, And III and Carina seem to be characterised
by more homogeneous stellar populations, with milder radial
dependences \citep{Harbeck01, Koch06, Koch07}.

A much closer look into specific dwarfs is allowed by large
spectroscopic datasets. For example, in Sculptor, Fornax and Sextans,
dedicated spectroscopic studies have allowed the identification of
independent stellar subpopulations, with chemo-dynamically distinct
properties {  \citep{Battaglia06, Battaglia08, Battaglia11, Walker11,
  Amorisco12b, Hendricks14}}. By providing a bridge between chemical and kinematical
properties of several hundreds of bright stars, chemo-dynamical
analyses yield important constraints on the complex processes that
govern the formation and evolution of such puny galaxies. For
instance, subpopulations with a higher average metallicity are more
centrally concentrated and kinematically colder. Whether the dwarf is
characterised exclusively by old stellar populations -- as in the case
of Sculptor \citep[see e.g.][]{deBoer12a} -- or whether multiple
populations of different ages are present at the same time -- as in
Fornax \citep[e.g.,][]{deBoer12b} -- the parallel {\it ordering} of
metallicity, characteristic scale of the stellar distribution and
kinematical state seems to be a fundamental outcome of the dSphs'
evolutionary histories.
 
The importance of Carina is that it provides one of the cleanest
examples of an episodic star formation (SF) history. This is evident
from its colour-magnitude diagram \citep{Tolstoy09, Stetson11}, which clearly
shows at least three different main sequence turn-offs. A number of
investigations have inferred the SF history of Carina using
photometric data \citep[e.g.,][]{Hurleykeller98, Rizzi03, Dolphin05}, with evidence
for at least two major SF episodes, one at old times ($>8$ Gyr
ago), a second at intermediate ages ($4-6\Gyr$ ago), and perhaps continuing into
even more recent activity ($2\Gyr$ ago). Using both wide-field photometry and spectroscopic
data, \citet{deBoer14b} find that about $60\%$ of the stars in Carina 
formed in the SF episode at intermediate ages. However, interestingly, 
as the oldest episode, also the intermediate one has enriched stars 
starting from low metallicities.

Given such a complex and episodic SF history, one would expect to
detect similarly clear population gradients and chemo-dynamical
differences in Carina's stellar population.  Based on deep multicolour
photometry, \citet{Harbeck01} find that red clump (RC) stars are indeed more
concentrated than the older horizontal branch stars. However, no
distinction in the spatial distribution of red and blue horizontal
branch stars can be found \citep[a gradient that is instead very clear
  in Sculptor,][]{Tolstoy04}.  More recently, \citet{Battaglia12} observe an age
gradient by selecting stellar populations of very different ages with
cuts in the colour magnitude diagram (CMD).  This confirms the existence of
a marked age gradient over a baseline that is essentially as long as
the age of the universe.

However, the presence of any gradient within the population of old red
giant stars is substantially less clear, as is the existence of a
corresponding metallicity gradient within the same population.  By
using a set of 437 radial velocity members, \citet{Koch06} measure a
very mild chemical gradient, with metal-poor stars having an only
slightly more extended spatial distribution. Analogously,
\citet{Walker11} could not identify any statistically significant
distinction (in either spatial distribution or kinematics) by looking
for two different subpopulations in a sample of more than 700 member
red giants (RGs).
 
Because of their shallow potential wells, the evolutionary histories
of dSphs are critically dependent on both properties before infall --
for instance virial mass and gas content -- and any subsequent
environmental factors, mainly driven by the details of their
orbits. In contrast to the wide orbits of Sculptor and Fornax,
Carina's proper motion \citep{Piatek03} suggests a pericenter of only a
few tens of $\kpc$ from the centre of the Milky Way, which makes it more
prone to tidal disturbances. Indeed, several investigations have
identified a component of `extratidal' stars around Carina
\citep[e.g.,][]{Majewski05, Munoz06}. Recent deep photometric studies
\citep{Battaglia12} suggest the presence of tenuous but extended
tails, and \citet{deBoer14b} has shown that these have compatible properties with Carina's metal poor
stars \citep[although see also][]{McMonigal14}. Interestingly, the perturbations arising from a tidal field
have been shown to be capable of systematically weakening chemical gradients,
by mixing any distinct stellar subpopulations {  \citep{Sales10,Pasetto11}}.

In this paper, we present a new dataset that we use in order to bring
further constraints on Carina's history, based on intermediate
resolution spectra of 956 stars. We describe how the analysis of the
spectra has been done in order to derive the atmospheric parameters of
the stars, namely the effective temperature (\teff), surface gravity
(\logg) and global metallicity (\meta), and how these parameters are
combined with high precision photometry to derive an age estimate for
the stars belonging to Carina.  We find evidence for the presence of
three distinct stellar subpopulations, and show the age-metallicity
diagram of Carina. Finally we comment on the possible effects of
external disturbances.

The paper is organised as follows: in Sect.~\ref{sec:data} we describe
the datasets used. Section~\ref{sec:parameters} explains how the
atmospheric parameters have been derived and the ages have been
estimated. Finally, Sect.~\ref{sec:description} shows the results and
Sect.~\ref{sec:conclusions} concludes.

\section{Presentation of the data}
\label{sec:data}

The available data include $1.3 \times 10^4$ medium resolution spectra
of 956 stars from the ESO-VLT large programmes, 180.B-0806, 084.B-0836
(P.I.: G.\,Gilmore) and our VISTA $JHK_s$ photometry. In addition, the
photometric data is complemented by the high-precision optical $UBVRI$
photometry of \citet{Stetson11}.

The VLT datasets extend over 5 years, with some multiple observations
of the same stars. The targets have been selected in order to sample
the red clump and the giant branch of Carina, and have been observed
using the LR8 setup of the FLAMES-GIRAFFE instrument
($820.6-940.0\nm$). For the spectra reduction and sky subtraction, we
refer the reader to \citet{Koch06}, since the same method has been
applied here too.  The radial velocities have been estimated using the
algorithm of \citet{Koposov11}, which delivers precise, accurate,
radial velocities from moderate resolution spectroscopy. The algorithm
works by fitting synthetic templates from \citet{Munari05} covering a
large range of stellar parameters ($-2.5 <$[Fe/H] $< 0.5$, 3000 $<$
\teff $<$ 80000, 1.5 $<$ \logg $< $5) to our spectra from each
exposure.

\begin{figure}
\begin{center}
\includegraphics[width=0.5\textwidth]{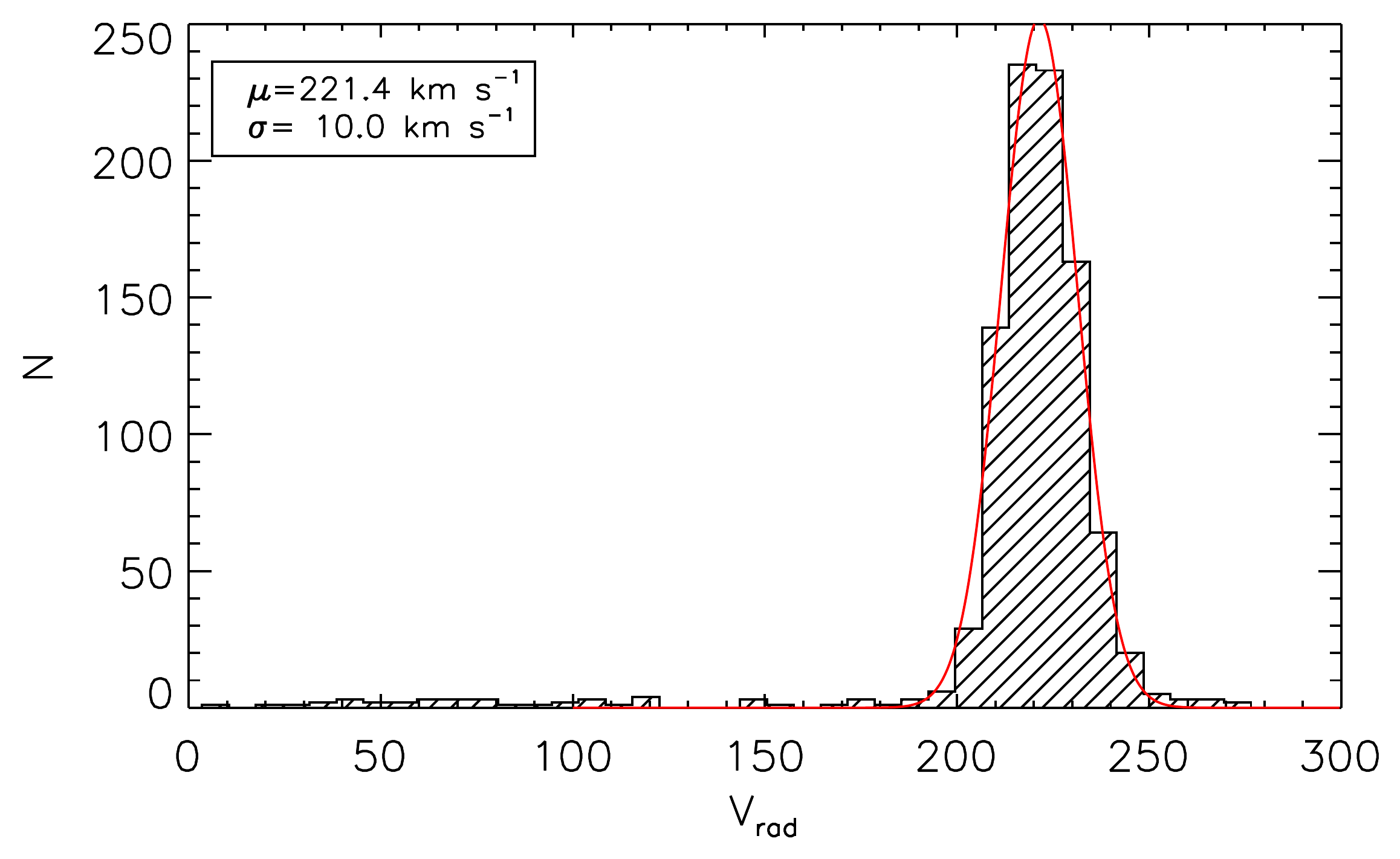} 
\end{center}
\caption{Histogram of the radial velocities for the total sample of
  stars observed at the LR8 setup (\totalLR~stars). The fitted
  Gaussian is plotted in red, and the fitting parameters are written
  on the top left corner of the figure. The mean value and dispersion
  of the radial velocity of Carina are in good agreement with the
  literature vaues.}
\label{Fig:Vrad_histogram}
\end{figure}

Once each exposure has been put in the rest frame, we have stacked the
multiple exposures corresponding to each star in order to increase the
signal-to-noise ratio (SNR) of the data. The final SNR spans values
from 2 to 50 per pixel.  Figure~\ref{Fig:Vrad_histogram} shows the
radial velocity distribution of the total sample of \totalLR\ stars
observed with the LR8 setup. From this, we can see that most of the
targeted stars peak at a common radial velocity, as expected. The
Gaussian fit of the histogram has a mean radial velocity of
$221.4\kms$ with a dispersion of $10\kms$, consistent with previous
determinations of the radial velocity of Carina star members ranging
from $220.4\kms$ to $224\kms$ and dispersions ranging from $6.8\kms$
to $11.7\kms$ \citep{Mateo98, Majewski05, Helmi06, Koch06, Fabrizio11,
  Lemasle12}.

\section{Extraction of the atmospheric parameters and selection of the Carina subsample}
\label{sec:parameters}

Once the spectra are stacked, we used an updated version of the
automatic parameterisation pipeline presented in \citet{Kordopatis11a}
to obtain the effective temperature, surface gravity and metallicity
of the targets.  The pipeline allows us to apply soft priors according
to the observed selection function, by removing from the solution
space combinations of parameters that are not expected to be found.
The method is based on a grid of synthetic spectra used during the
learning phase of the algorithm.  Only the wavelength regions
$840-877.5\nm$ and $880.1-882.0\nm$ are selected for the
parameterisation, the discontinuity being introduced to avoid strong
contamination by telluric lines {  and to keep within our wavelength range the MgI line at 8807\AA, which is known to be sensitive to surface gravity variations} \citep[see][]{Kordopatis11b}.
{  Furthermore, the cores of the Calcium triplet lines are removed from the wavelength region (two pixels for the first line, and three for the other two lines, corresponding to  0.8  and  1.2\AA,  respectively), in order to avoid a mis-match between the synthetic spectra and the observations, due to non local thermodynamical equilibrium effects.}

The learning grid has a constant metallicity step of $0.25\dex$, and
spans effective temperatures from $[3000-8000]$\,K, surface gravities
from $[0-5]$($\cgs$ units) and metallicities from
$[-5.0,+1.0]$. Finally, the $\alpha-$enhancement of the considered
templates is not a free parameter, but it varies with metallicity
($[\alpha/$Fe$]=-0.4\times \feh$ in the range $-1\leq \feh \leq0$).
{  We note that this adopted $\alpha$-enhancement has no consequences for the determination of \meta\, in the case where a star does not follow the same trend \citep[as is expected to be the case for Carina dSph stars, see][]{Venn12}. In such a case the measurement of $\meta$\, will still be sound, but \meta$\ne$[Fe/H] \citep[see][for further details]{Kordopatis13b}.}

We used the calibration relation established for RAVE DR4 \citep[][see
  also \citealt{Kordopatis15a}]{Kordopatis13b}, which employs the same
grid of synthetic spectra on a very similar wavelength range, to
correct the metallicities of the pipeline. The calibration is a simple
low-order polynomial of two variables, the surface gravity and the
metallicity itself, and roughly corrects the metallicity of the giants
by $\sim0.3\dex$ and the one of the dwarfs by $\sim 0.1\dex$. The
adopted relation allowing to obtain the calibrated metallicity,
\meta$_c$, from the one derived from the pipeline, \meta$_p$, is:
 \begin{small}
\begin{equation}
\begin{split}
\meta_{c} - \meta_{p}=-0.076 - 0.006 \times \log g + 0.003\times\log^2g \\
- 0.021\times \meta_{p}\times\log g 
+0.582\times \meta_{p}+0.205\times \meta_{p}^2.
\end{split}
\end{equation}
\end{small}
In what follows, only the corrected metallicities will be used, and
therefore we will note the calibrated metallicities simply as [M/H].

\begin{figure}
\begin{center}
\includegraphics[width=0.5\textwidth]{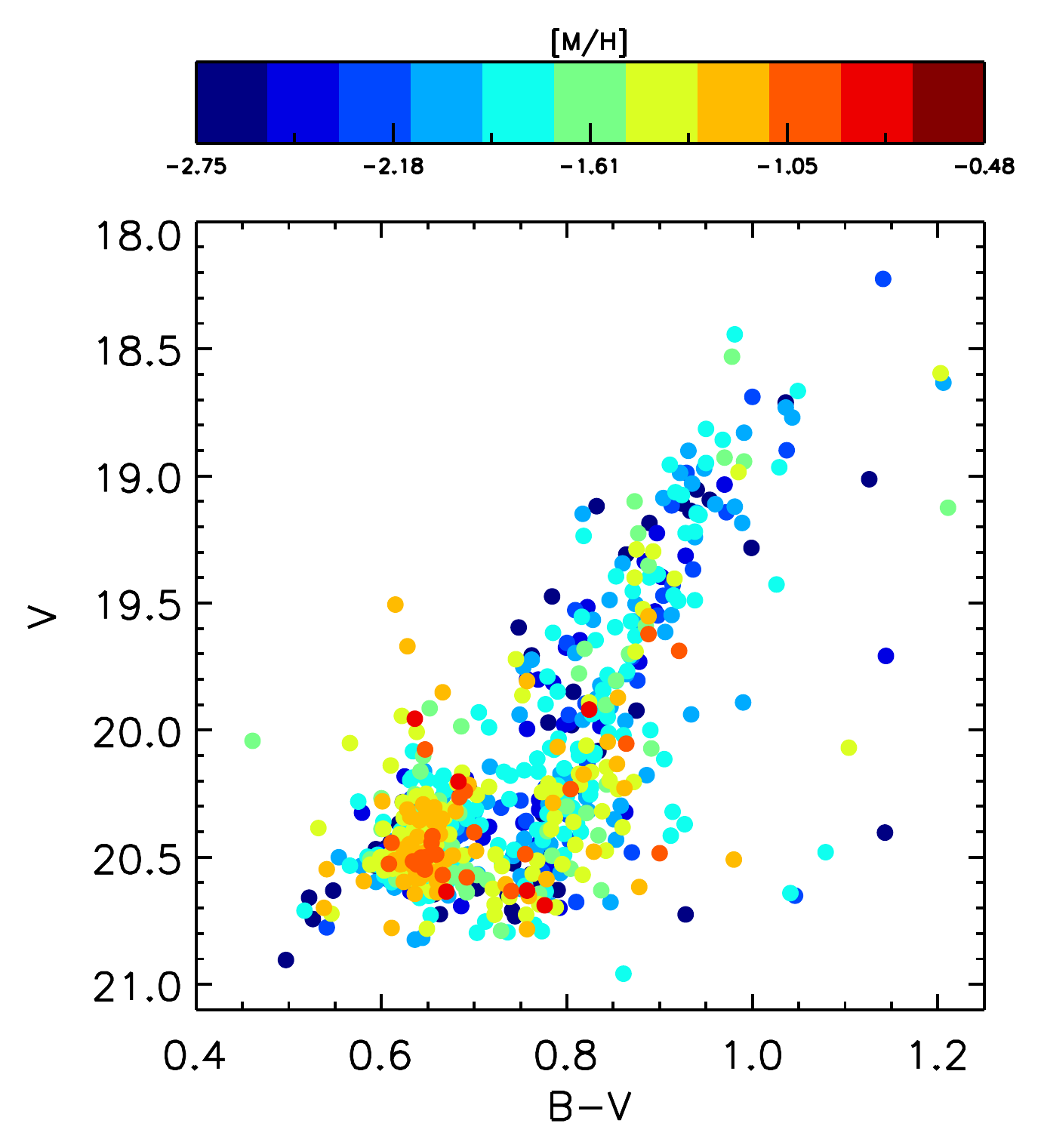} 
\end{center}
\caption{Colour-magnitude diagram of the final adopted Carina sample
  (\totalCarina\ members). The RC is at $(B-V)<0.7$ and $V>20$,
  whereas the RGB contains all the stars with $(B-V)>0.7$ up to
  $V=18$. }
  \label{Fig:CMD}
\end{figure}

 As noted at the beginning of the Section, soft priors have been
 imposed on the expected results, by removing some parameter
 combinations from the solution space. For the adopted priors, we
 imposed:
 \begin{itemize}
 \item
  an effective temperature between $4000-6500$~K,
  \item
  a surface gravity lower than $3.75$, 
  \item
  a metallicity range between $-5$ and $+1$ (i.e. all the available metallicity range of the templates). 
  \end{itemize}
The temperature range can be justified by the colour selection of
Carina's candidates ($0.5<B-V<1.2$, see Fig.\,\ref{Fig:CMD}).  As fas as the assumption on the
surface gravity is concerned, the reason relies on our {\it a priori }
knowledge of the properties of the observed stars.  Indeed, given the
distance modulus of Carina \citep[$m-M\sim20.1$, e.g.][]{Dallora03},
all of Carina's main sequence stars should be outside the observed
magnitude range.  In other words, all the stars belonging to Carina
are expected to be either red clump stars or on the red giant
branch, hence with a \logg\ lower than 3.5.

 \begin{figure}
\begin{center}
\includegraphics[width=0.5\textwidth]{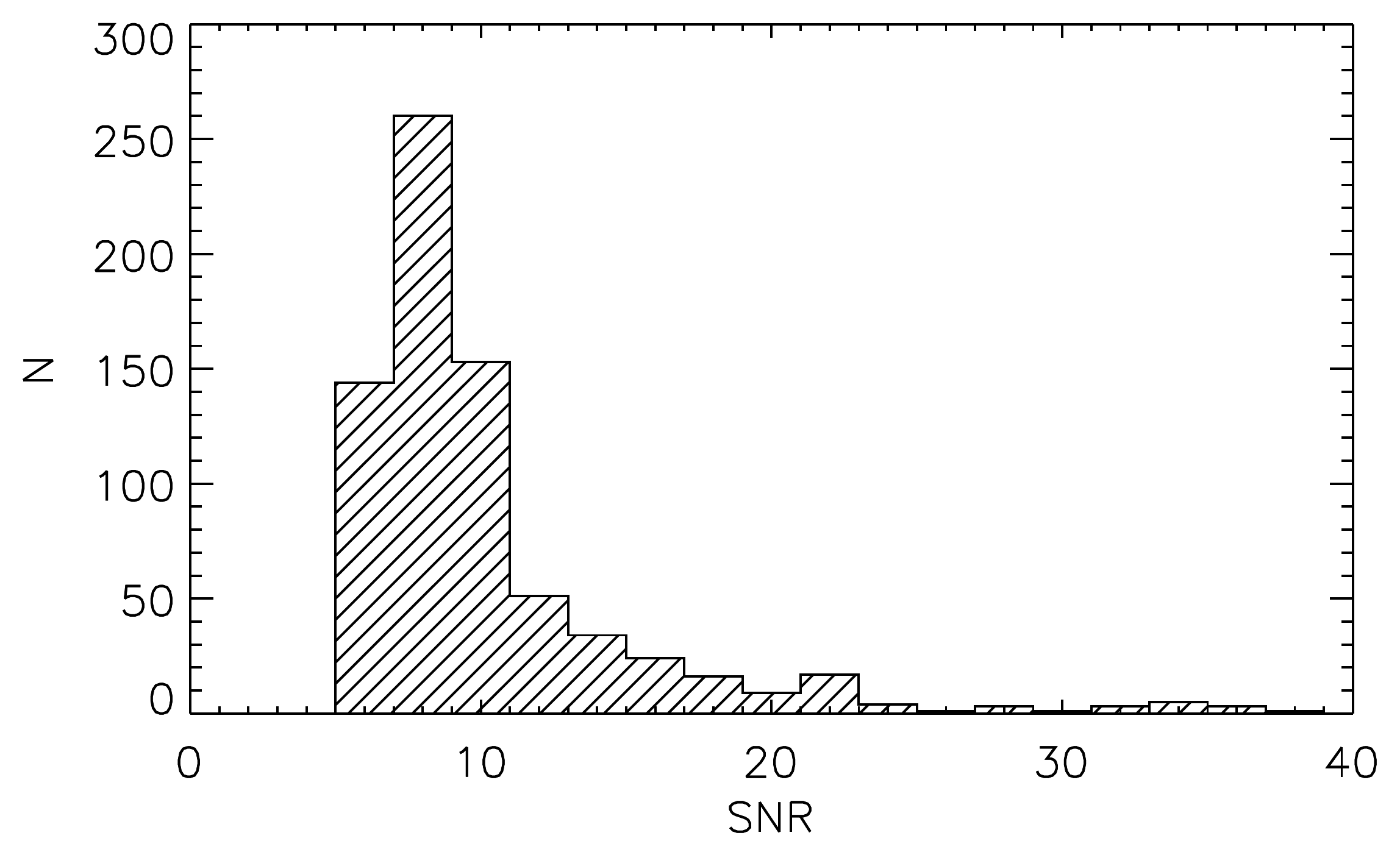} 
\end{center}
\caption{Histogram of the mean signal-to-noise ratio per pixel of the
  spectra of the final Carina sample that is considered for the
  analysis (\totalCarina\ members).}
\label{Fig:SNR_histogram}
\end{figure}

\begin{figure}
\begin{center}
\includegraphics[width=0.5\textwidth]{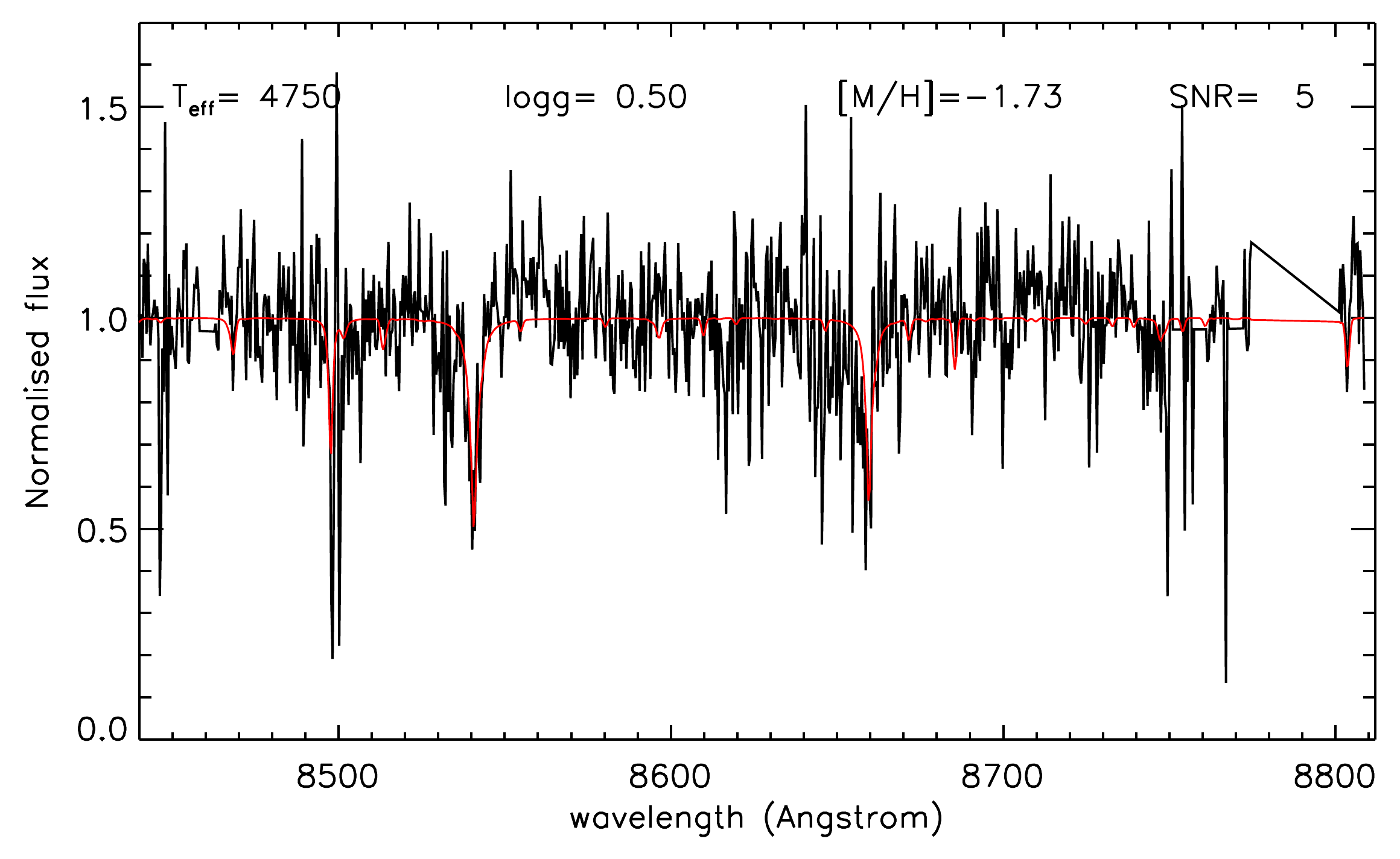}\\
 \includegraphics[width=0.5\textwidth]{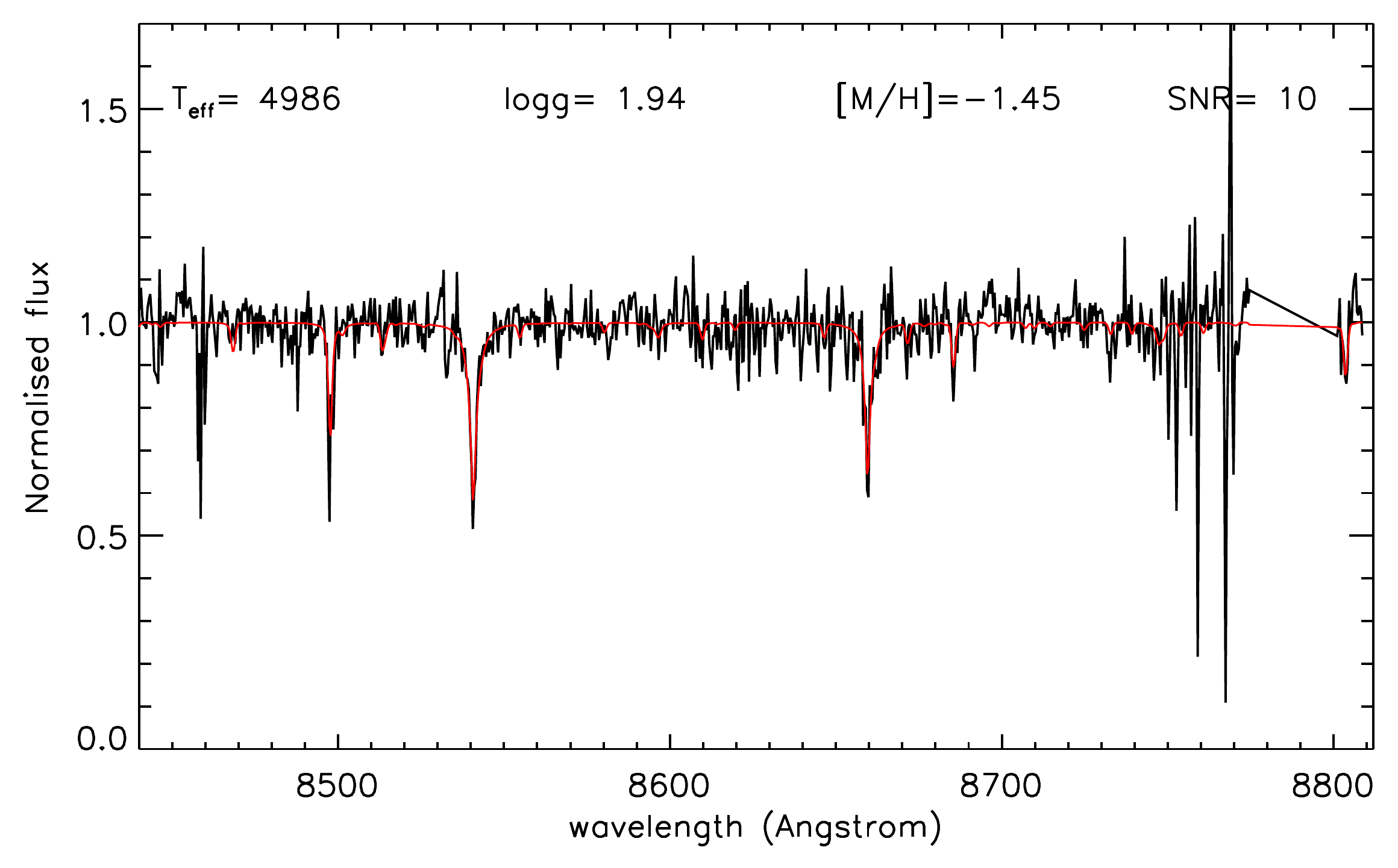}\\
 \includegraphics[width=0.5\textwidth]{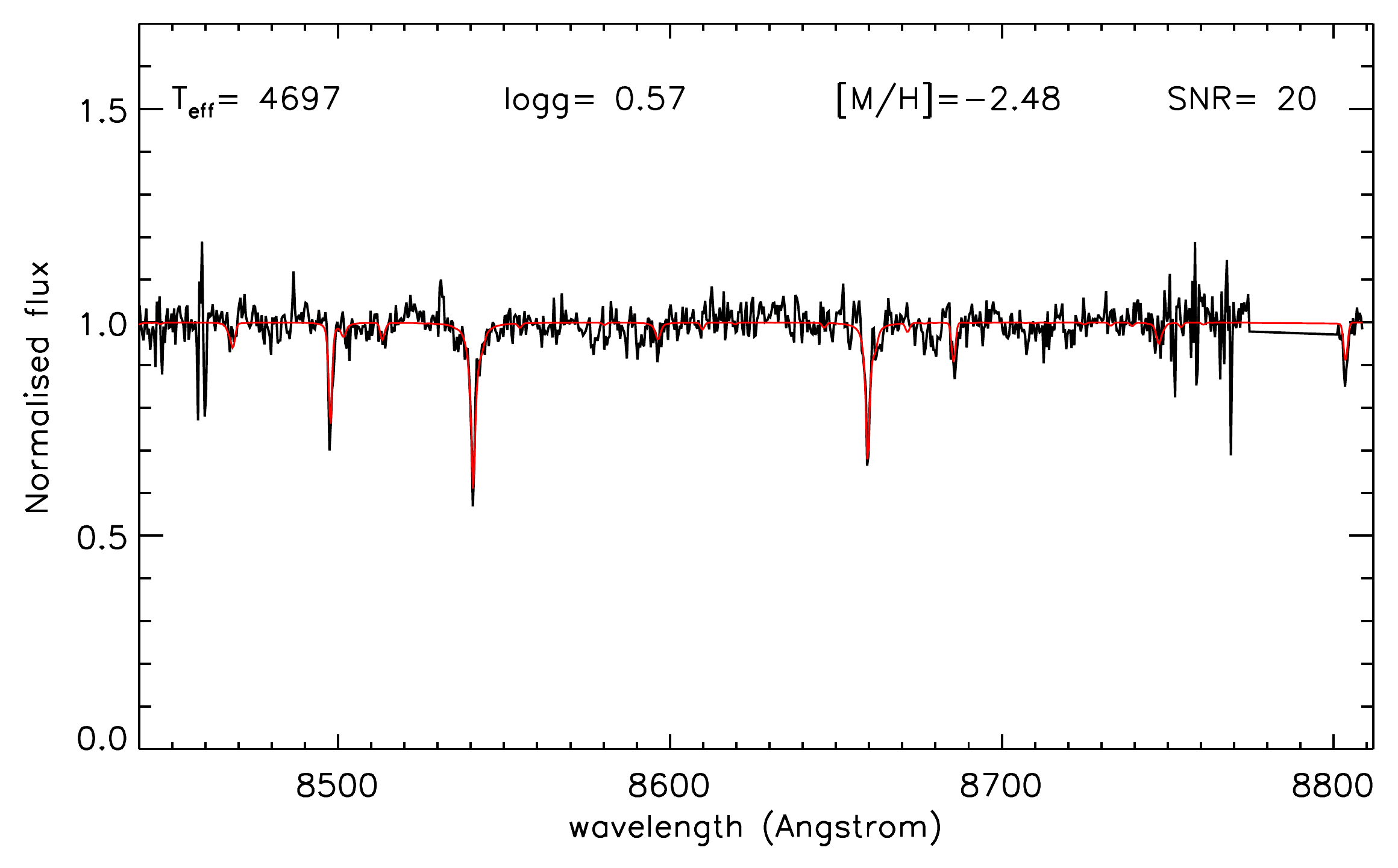}\\
\end{center}
\caption{{  Observed (black) and fitted spectra (red) corresponding to the synthetic spectrum having the parameters of the stars for three stars having SNR$\sim5, 10,20$ ~pixel$^{-1}$. The wavelength range has been truncated at the blue end by $40$\AA\, to make the plots clearer to visualise.}}
\label{Fig:Obs_spectra}
\end{figure}

That said, one must still understand the effect of removing regions of
the solution space on the parameterisation of foreground stars that
might have the same radial velocity as Carina, and contaminate our
sample (the risk here being to mis-parameterise a dwarf star as a
giant). A first statement that can be made is that given the radial
velocity of Carina ($221\kms$), foreground stars having a similar
radial velocity are expected to be mainly halo stars, and therefore
giants.  For the few foreground stars that might still have
\logg$>$3.75, the algorithm will always match, by design, the
observations with the closest template (the latter being the one
having the closest parameters as the true spectrum). This implies that
the derived parameters of a dwarf star will therefore be at the
boundaries of the grid and easily identifiable. In order to make sure
that such contaminators are excluded from our analysis, we further
decided to discard all the stars for which the surface gravity is
greater than $3.25$, as well as the stars with a spectra having a SNR
lower than 5\,pixel$^{-1}$, because of the large uncertainty in the
derived parameters. Figure~\ref{Fig:SNR_histogram} shows the SNR
histogram of the Carina sample that is considered in what follows. The
bulk of the spectra have SNR$\sim$10 ~pixel$^{-1}$, with some of them
having up to SNR$\sim$40 ~pixel$^{-1}$. 
{  A selection of three spectra, at SNR$\sim5, 10,20$ ~pixel$^{-1}$, with their best-fit solution are shown in Fig.~\ref{Fig:Obs_spectra}.}

 \begin{figure}
\begin{center}
\includegraphics[width=0.5\textwidth]{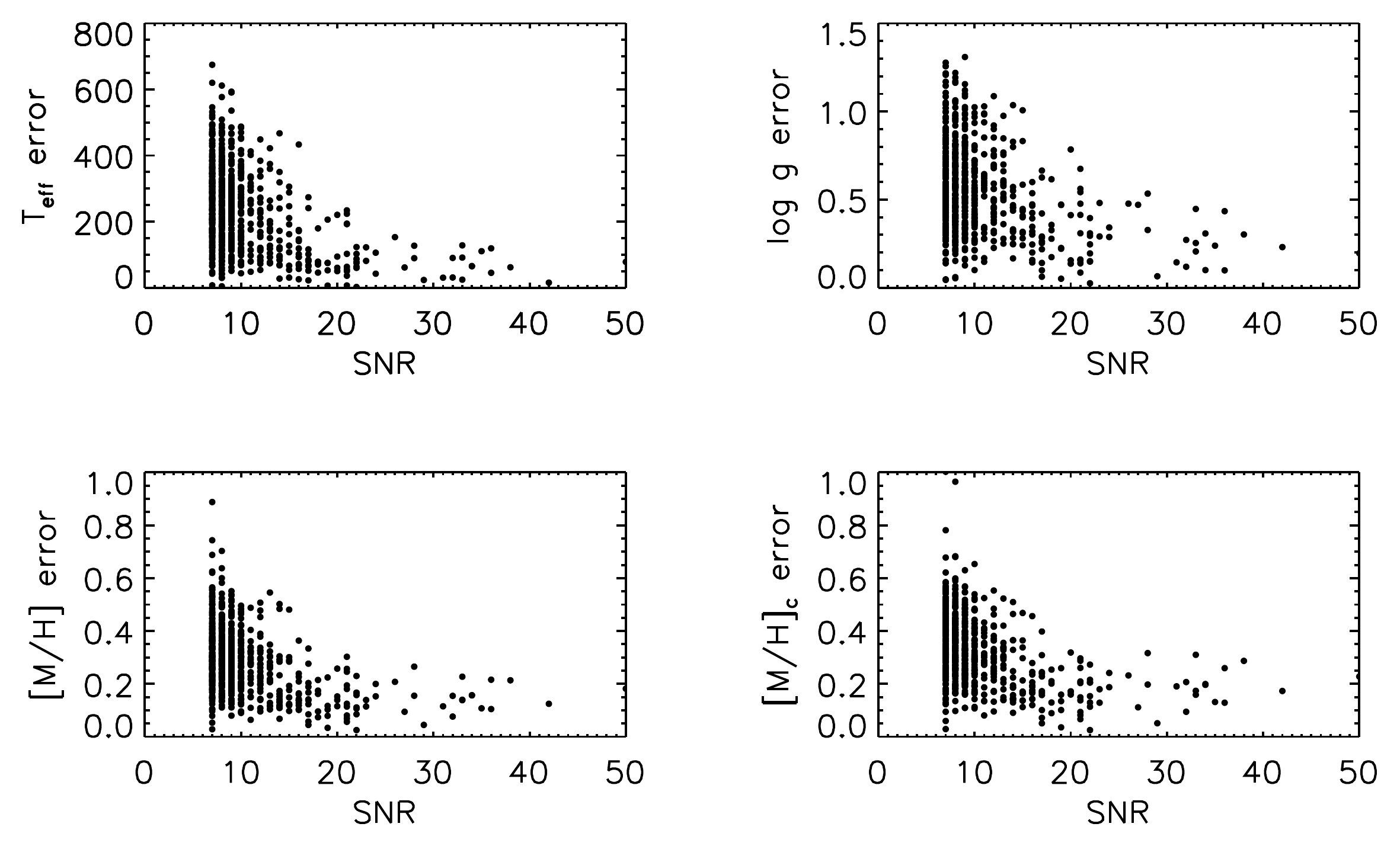} 
\end{center}
\caption{{  Errors on the effective temperature, surface gravity, uncalibrated metallicity and calibrated metallicity as a function of signal-to-noise, for the sample of stars that we have identified as Carina members and with SNR greater than 5\,pixel$^{-1}$. }}
\label{Fig:SNR_errors}
\end{figure}

Typical uncertainties on the parameters were obtained using the error
spectrum of each target, producing 10 Monte-Carlo realisations of
observed spectra and re-deriving the parameters. For \teff, \logg\ and
[M/H], the median uncertainties are of the order of 226\K, 0.48\dex\,
and 0.29\dex, respectively, {  with the errors being the largest for the lowest SNR values, as expected (see Fig.~\ref{Fig:SNR_errors})}.  

Figure~\ref{Fig:CMD} shows the CMD for
Carina for our adopted sample, colour-coded according to the
metallicity of the stars.  We find that the mean metallicity for the
whole sample is $\meta\approx-1.78$, with a large span in the derived
values, ranging from $\sim-4$ to $\sim-0.5$. This result is in very good
agreement with previous studies \citep[see for example][]{Koch06}.

\subsection{Age determination of the red giant branch stars}
\label{sec:ages}

A rough estimate of the age of the stars can be obtained by comparing the colour,
magnitudes and atmospheric parameters of the stars with theoretical
isochrones.  Following \citet{Kordopatis11b}, we constructed a library
of isochrones with a constant step in age ($0.5\Gyr$) and metallicity
($0.1\dex$). The step in metallicity has been chosen in order to be
smaller than the typical error on this parameter.  The isochrones have
been computed using the online
interpolator\footnote{\url{http://stev.oapd.inaf.it/cgi-bin/cmd}} of
the Padova group, based on the \citet{Marigo08} sets, for a
metallicity range of $[-2.1;0.1]$.
 Since the isochrones do not reach as low metallicities as the metal-poorest stars in our sample, we do not attempt to derive ages for those stars that have a metallicity that does not reach $-2.1$ within $2\sigma_{\tiny \meta}$,  where $\sigma_{\tiny \meta}$ is the uncertainty on the derived metallicity.

The expected age, $\bar{a}$, of a star with parameters $\hat \theta_k$ ($k \equiv \meta$, $V$, $B-V$, \teff, \logg), has been obtained as follows. First, we select the set of isochrones within \meta$\pm \sigma_{\tiny \meta}$. Then, we assign for each point $i$ on the selected isochrones a Gaussian weight $W_i$ , which depends on the distance between the points on the isochrones and the measured observables or derived parameters. In practice,  $W_i$ is computed as:
 \begin{equation}
W_i= \mathrm{exp}\left(-\sum_k \frac{(\theta_{i,k} - \hat \theta_k)^2}{2\sigma^2_{\hat \theta_k}}\right)
\label{eqn:weight_isochrones}
\end{equation}
where $\theta_{i,k}$ corresponds to the considered parameters of the isochrones and
$\sigma_{\hat \theta_k}$ to the associated uncertainties of the
measurements. It is worth mentioning that we did not include to this
weight any additional multiplicative factor proportional to the mass
of the stars, as suggested by \citet[][see also
  \citealt{Kordopatis11b}]{Zwitter10} because this factor is useful
only for surveys having a mixture of dwarfs and giants. Indeed, this
factor defined as the stellar mass difference between two adjacent
points on the isochrones, is introduced in order to give additional
weight to the likelihood of observing a dwarf, because they are
characterised by slower evolutionary phases. Since our survey includes
only giant stars, this factor has been ignored.

The expected age $\bar{a}$ of a given star is  then obtained by computing the weighted mean: 
\begin{equation}
\bar{a}=\frac{\sum_{i} W_{i} \cdot   a_i}{\sum_{i} W_i },
\end{equation}
where $a_i$ are the associated ages of the points on the isochrones. The associated error of the expected age is obtained by: 
\begin{equation}
\sigma(\bar{a})=\sqrt{\frac{\sum_{i} W_i \cdot   [\bar{a} - a_i ]^2}{\sum_{i} W_i }}.
\end{equation}

\begin{figure}
\begin{center}
\includegraphics[width=0.5\textwidth]{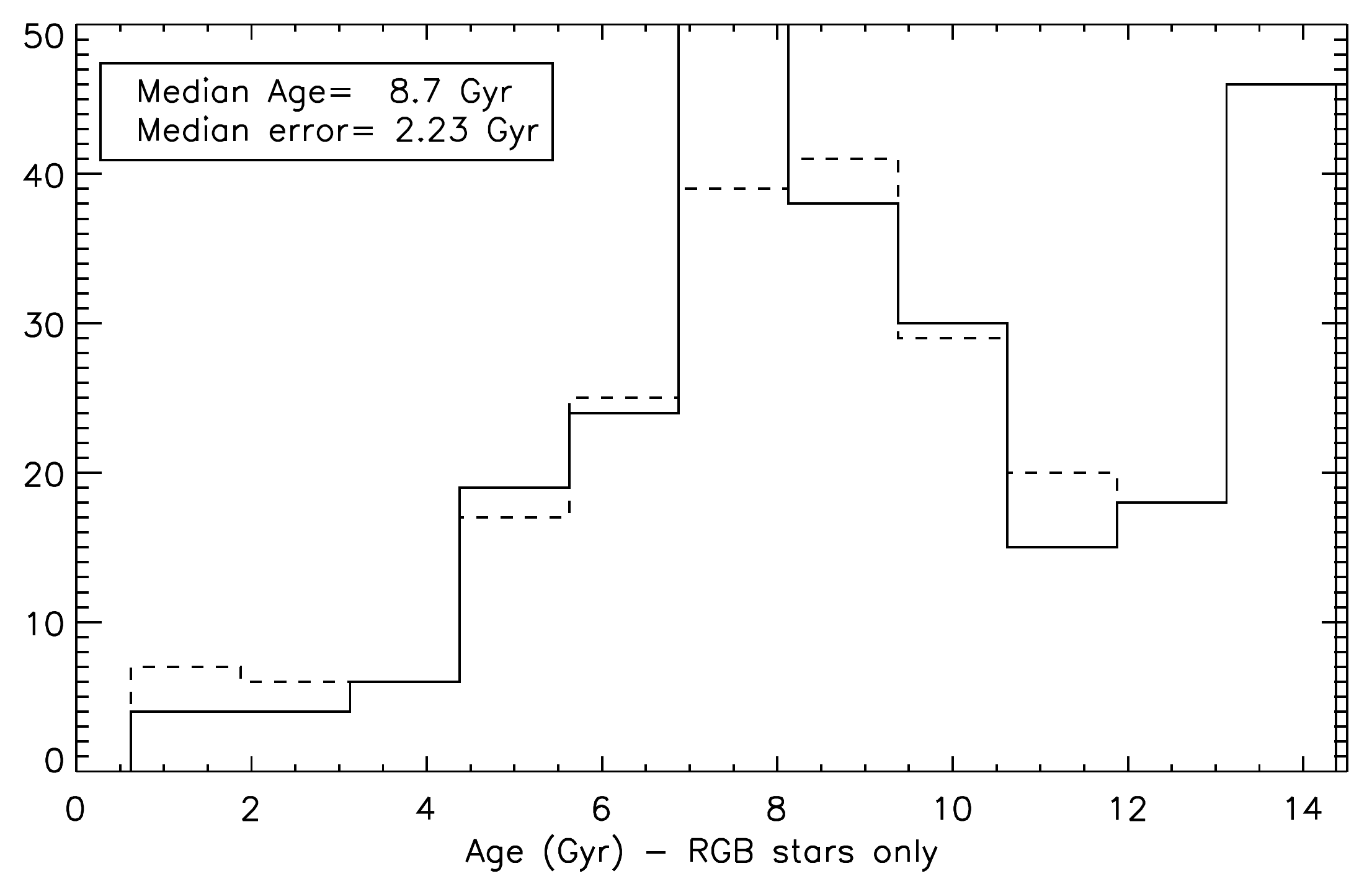} 
\end{center}
\caption{Age distribution of the RGB stars of Carina. The dotted
  (plain) line corresponds to age estimations with (without) taking into
  account the information on the effective temperature and gravity. }
\label{Fig:Ages}
\end{figure}

We assumed a distance modulus for Carina of $m-M=20.1$ and a line-of-sight $E(B-V)=0.03$~mag, as
estimated by, e.g. \citet{Dallora03} and \citet{Karczmarek15}\footnote{ {We note that our age estimation is rather robust to the adopted distance modulus. Indeed, assuming for example $m-M=20.3$\,mag \citep[see][for a discussion of possible values of the distance modulus]{Vandenberg15}, leads to a median change in the final ages of $0.3\Gyr$ with a $\sigma$ of $1.6\Gyr$.}}. The
considered age-range of the isochrones has been set to be between
$1\Gyr$ and $13.7\Gyr$.  We have tested two different configurations:
one where only $V$, $(B-V)$ and the metallicity are taken into
account, and one where we additionally consider the information
related to the effective temperatures and surface gravities.
Figure~\ref{Fig:Ages} shows the differences in the age estimations
according to these two approaches for the RGB stars only (for the RC stars, see the discussion below). We can see that our sample can
be separated at least into two populations, one old ($\sim 13\Gyr$), and one of
intermediate age having a peak at $\sim7.5\Gyr$ and with a much
broader age range. This result is in good agreement with, for
example \citet{Stetson11}, who fit the turn-off stars in the CMD and
estimate that Carina has at least two populations: one of $12\Gyr$ and
one of $4-6\Gyr$, or with Norris et al. (in prep) who are completing a similar age-metallicity analysis using high-resolution spectra of giant stars.
  In particular, we find that this intermediate age
population has a tail extending to young ages, down to $1\Gyr$. Taking
into account the atmospheric parameters in
Eq.~\ref{eqn:weight_isochrones} does not change the age distributions drastically,
as can be seen from the dotted (with \teff, \logg) and plain
(without \teff, \logg) histograms of Fig.~\ref{Fig:Ages}. This
similarity in the age estimations shows that our effective
temperatures, gravities and metallicities are consistent with the CMD
of Carina. However, since the atmospheric parameters can have large covariances, and that the ages are not fundamentally changed when taking into account the \teff\ and the \logg, in what follows, we have considered only the ages computed
without taking into account the atmospheric parameters.
 
We note that for 10 stars, no age estimation was possible. The reason
is that they are too far from the isochrones, either because they are
foreground contaminators, binary stars, or stars for which the
measurements (metallicity and/or photometry) are of poor quality and
with underestimated errors. 

\subsection{Age determination of the red clump-region stars}

The RC  stars are massive stars which have
passed through the explosive ignition of the Helium at the RGB tip and
are now burning the helium core.  Once all the helium in the core is
exhausted, then the stars begin their AGB phase.  Their absolute
magnitude mildly depends on the metallicity and their age, with the
oldest and most metal-poor ones being also the faintest.

The determination of the age of the stars in the RC region of the CMD ($B-V\leq 0.7$) by the projection on the
isochrones is generally difficult to obtain because all the isochrones
pass through that region. 
A first run of our pipeline on
the stars in the RC-region has shown that the estimated ages of the bulk of these stars is
similar to the intermediate age population of the RGB ($\sim 7.5\Gyr$
for the RGB, $\sim 6\Gyr$ for the RC-region), as expected from the observing
biases (at these magnitudes, we do not see the oldest RC stars with
still fainter magnitudes).  Nevertheless, we find a non-negligible
fraction of stars with ages lower than $4\Gyr$. 

{  Since they do not have a counter-part (in terms of star-counts) on the RGB, 
it seems clear that the ages of the stars in the RC-region having low metallicities should have much older ages than what is derived (therefore likely being the oldest stars that have left the Horizontal Branch, and are now on the AGB}. If these stars do follow the derived age-metallicity relation of the RGB stars, then their estimated ages should be $\sim13\Gyr$. Given these facts we decide, in what follows, not to take into account the ages of the RC-region stars in our analysis.

 \section{Chemodynamical separation of the stellar  populations} 
 \label{sec:description} 
 
In order to identify and separate any chemodynamic stellar
subpopulations, we use the maximum likelihood technique presented by
\citet{Walker11} and later developed in \citet{Amorisco12b}. 
We model the spectroscopic target as the combination of multiple subpopulations with different intrinsic properties, including metallicity, kinematics and spatial distribution. 
{  The crucial advances in this technique are that:
\begin{enumerate}
\item
  all available information can be used at the same time, improving
  the quality of the population division with respect to a method that only uses one dimension at a time (for example metallicity or kinematics, separately); 
 \item
  as a result of the population division, each single star is tagged with its membership probability to all identified subpopulations, so that with respect to the cross-contamination due to hard cuts, such a mixture model can better disentangle the properties of any different component; 
 \item
  the number of independent subpopulations needed to best describe the data can be determined objectively, by comparing the gain in likelihood due to the increased number of populations to the growth in the number of free parameters of the model; 
 \item
  any selection function can be explicitly taken into account. 
 \end{enumerate}
Details on the technique can be found in \citet{Walker11} and
\citet{Amorisco14b}.}  The mapping of the chemodynamical subpopulations
onto the space of stellar ages is presented in
Sect.~\ref{sec:age_decomposition}.

\begin{table*}
 \centering
  \caption{Three distinct chemo-dynamical subpopulations in the Carina dSph: the structural parameters.}
  \begin{tabular}{@{}lcccccc@{}}
  \hline
   Subpop     &    $\langle[{\rm M/H}]\rangle$ & StD$([{\rm M/H}])$ & $R_{\rm h}/$pc & $e$ & $\langle \sigma_{LOS}\rangle/$kms$^{-1}$ & $f$\\
\hline
RGB Metal Poor (MP) & $-2.4\pm0.1$ & $0.21\pm0.08$ & $500^{+150}_{-75}$ & $0.5\pm0.15$ & $10.4\pm1$ & $0.21\pm0.06$\\ 
RGB Intermed. Met. (IM) & $-1.84\pm0.05$ & $0.10\pm0.07$ & $185\pm20$ & $\lesssim0.2$ & $7.6\pm0.5$ & $0.59\pm0.04$\\ 
RGB Metal Rich (MR) & $-1.0\pm0.1$ & $0.23\pm0.07$ & $400^{+100}_{-50}$ & $0.45^{+0.15}_{-0.1}$ & $8.5\pm0.8$ & $0.20\pm0.05$\\ 
\hline
Red Clump & $-1.50\pm0.05$ & $0.50\pm0.05$ & $225 \pm 20$ & $0.30 \pm0.06 $ & $8.6 \pm 0.3$ &  -- \\ 
 \hline
\end{tabular}
\label{tab:pop_numbers}
\end{table*}

\subsection{Three distinct red giant branch subpopulations}

\begin{figure*}
\centering
\hspace{-.5cm}
\includegraphics[width=.5\textwidth]{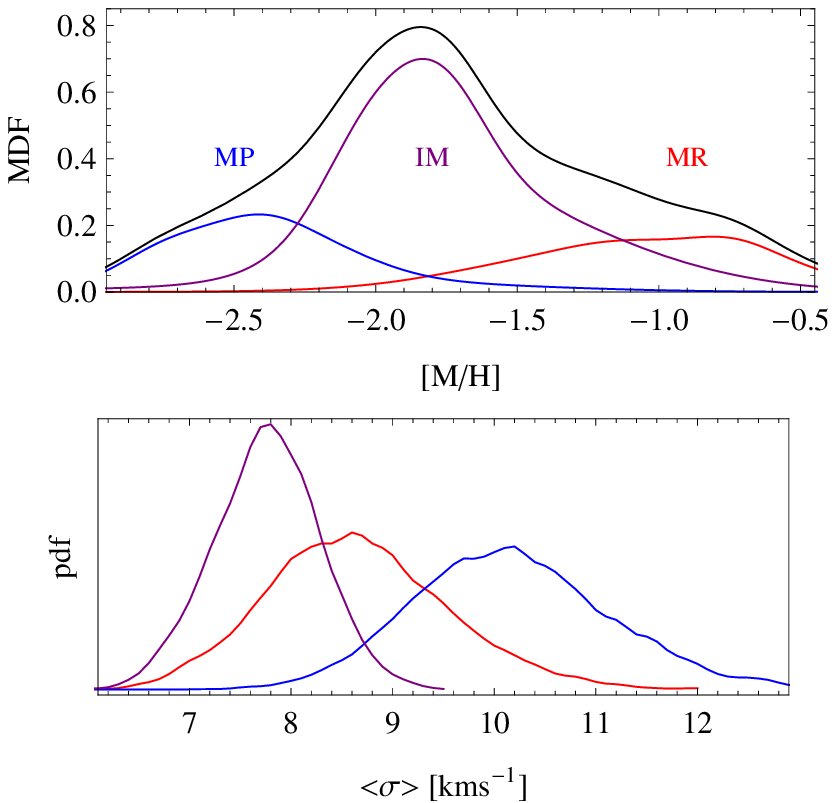}
\includegraphics[width=.4\textwidth]{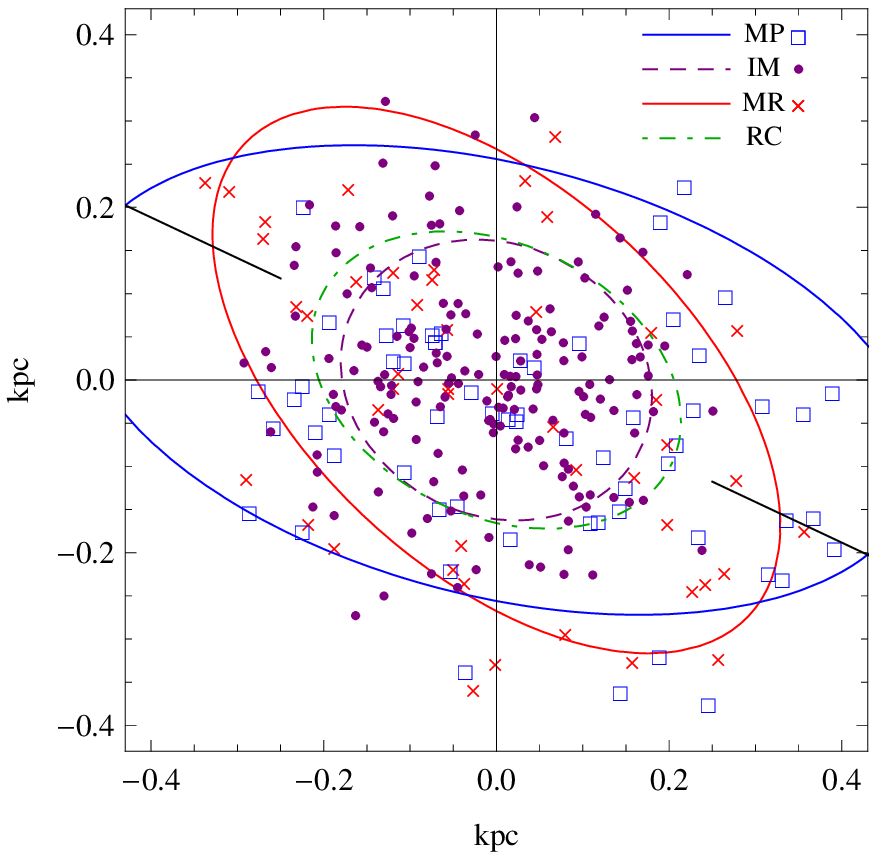}
\caption{Three distinct subpopulations. 
  Left upper panel: metallicity distribution function decomposed in three populations. 
   Black: the MDF of the entire RGB sample. Blue, purple and red, respectively the 
   contributions of metal-poor, intermediate-metallicity and metal-rich populations
   to the global RGB MDF. Note that distribution appear broader than the measured 
   intrinsic spreads because of uncertainties on the discrete metallicity measurements.
   Left lower panel: the inference for the average velocity dispersion of the three RGB subpopulations.
  Right panel: the spatial distribution of the {  stars having more than 75 percent probability of being members} 
  (metal rich population
  in red, intermediate metallicity in purple, metal poor in blue). Coloured ellipses display the best fitting half-light radii
  for the each subpopulation; additionally the red clump population is in green and black lines display the position angle 
  of the photometry and direction of Carina's tidal tails. }
\label{pops}
\end{figure*}

We first restrict ourselves to the subset of high-probability Carina members
belonging to the RGB ($B-V>0.7$, 400 members)
and investigate any chemo-dynamical sub-divisions. We find that a
two-populations model is preferred over one with a single population,
but a model including three-populations represents the best description of the
data (the probability of obtaining an analogous gain in the likelihood
function by pure chance is negligible, despite the additional degrees of freedom
of the three-populations model). Table\,\ref{tab:pop_numbers} collects the properties of the three 
identified subpopulations.

We find that chemistry is the main driver of the division: we identify
a metal poor (MP) population, with average metallicity of about
$\langle[{\rm M/H}]\rangle\approx-2.4$, a population of intermediate
metallicity (IM, $\langle[{\rm M/H}]\rangle\approx-1.8$) and a metal
richer (MR) population, with $\langle[{\rm M/H}]\rangle\approx-1$.
The upper-left panel in Fig.~\ref{pops} displays the metallicity
distribution function of our sample of red giant stars and illustrates
the division in subpopulations.  The bulk of Carina's red giants
belong to the population of intermediate metallicity, collecting a
fraction $f_{\rm IM}\approx60\%$ of the total RGB population,
similarly to what found by \citet{deBoer14b}.  Note that this plot
includes both (i) a convolution with the individual observational
uncertainty of each metallicity measurement and (ii) the partial
weighing of each star with its membership probability to each
subpopulation. The first is responsible for the larger metallicity
spread of each population (with respect to the intrinsic spread listed
in Table~\ref{tab:pop_numbers}), while the second is responsible for
the skewness and overlap between the different populations {   due to those stars that have similar probabilities of belonging to either population (caused by both observational uncertainties in metallicity and by the similarity between the recovered kinematical properties of the populations)}.

The right panel of Fig.~\ref{pops} shows the spatial distribution of
the three subpopulations, with symbols identifying the high
probability members {  ($p\geq0.75$)}.  Each ellipse corresponds to the best fitting
elliptical half-light radius $R_{\rm h}$, with its ellipticity and
position angle.  
{  The properties of the spatial distributions of the different populations are obtained by assuming a parametric functional form (Plummer profile) and by fitting for the spatial distribution of discrete spectroscopic targets, following \citet{Walker11}, but also allowing for a non zero ellipticity.}

The MP population is considerably elongated and
extended. In fact, we are not entirely able to measure its half-light
radius, as we find that its distribution in the radial range probed by
our data is almost flat. We find that MP stars are approximately
aligned with Carina's tidal tails, as measured by
\citet{Battaglia12}. The MR population is somewhat less extended,
although also quite spread out over our radial coverage.  On the
contrary, the IM population is substantially more compact and
centrally peaked, with a well defined half-light radius and a
decreasing number counts profile.

\begin{figure}
\begin{center}
\includegraphics[width=0.4\textwidth]{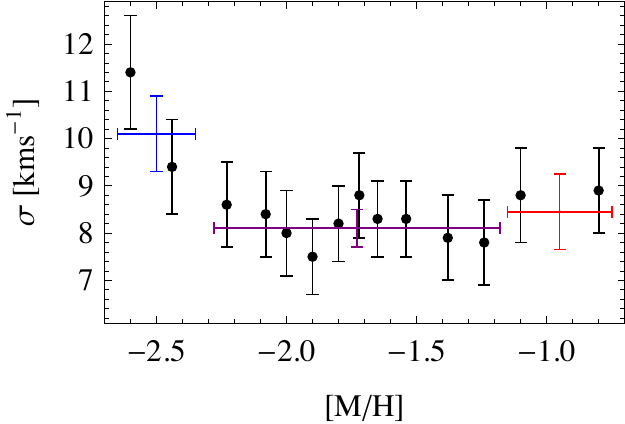} 
\end{center}
\caption{ Projected velocity dispersion versus metallicity, obtained by simply binning the spectroscopic data for the RGBs. Black points are obtained using subsets of  50 RGBs each (successive bins shift by 25 stars at a time, so that not all data points are independent). All measurements are obtained through a maximum likelihood method, and the vertical error bars show $1\sigma$ uncertainties. The wider and coloured error bars group RGB stars into wider bins, and are fully independent from each other (respectively, the three bins contain 75, 250, 75 RGBs).}
  \label{Fig:Simple_decompositions}
\end{figure}

In contrast with all other previously studied dSphs, we find that the MR
population is more extended than the IM population, and, accordingly, its
characteristic velocity dispersion is most likely larger, as shown in the 
lower-left panel of Fig.~\ref{pops}. 
{  This inversion can be captured even with hard cuts in metallicity, i.e. without the use of the likelihood division method. Although such a binning makes the signal weaker due to the large errors in metallicities, the trends shown in Fig.~\ref{Fig:Simple_decompositions} indicate, indeed, that the high metallicity tail is at least as kinematically hot than the bulk of the stars in the intermediate-metallicity bin. The same cuts can be used to show that the spatial distribution of the stars in the intermediate-metallicity bin is  more compact than that of the high-metallicity tail, proving that our result is robust.}

This represents the first
case in which it is not possible to identify a complete parallel {\it ordering}
of metallicity, characteristic scale of the stellar distribution and
kinematical state. The MR population is more extended than the
IM population, and, at the same time, at least as kinematically hot.
In turn, this may provide a justifications to previous measurements of a very 
limited global chemical gradient in both \citet{Koch06} and \citet{Walker11}.

\subsection{Comparison with the Red Clump population}

We have decided to carry out a separate analysis of the RC
population, keeping it distinct from the RGB population.
We do not try to separate subpopulations based on chemistry within the RC, and 
instead only measure global properties of their spatial distribution,
kinematics and chemistry.

We use more than 400 line-of-sight velocity measurements and
associated spatial position for RC stars that belong to Carina with
high probability. As listed in Table~\ref{tab:pop_numbers}, we find that the RC population
bears significant similarities with the population of intermediate
metallicity in the red giant branch.  Its half light radius is
comparable with the half light radius of the photometry and so are its
ellipticity and position angle (as shown in the right panel of
Fig.~\ref{pops}). Note that, being on average slightly younger, it is likely
that the RC population probes a combination of the IM and MR red giant
subpopulations. Therefore, the fact that the IM population is even
more concentrated than the RC population is a confirmation of the
inversion described in the previous section.  The same reasoning
applies to the kinematics of the RC population, whose velocity
dispersion is slightly hotter than the the one of the IM population.

\subsection{Age decomposition of the red giant branch stars}
\label{sec:age_decomposition}

\begin{figure}
\centering
\includegraphics[width=.45\textwidth]{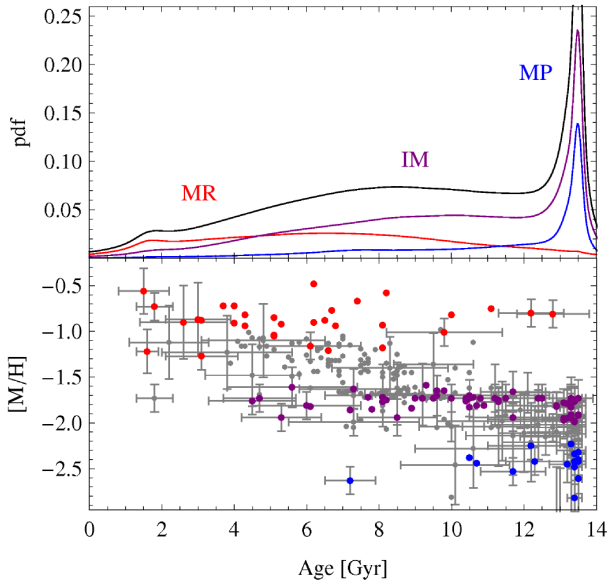}
\caption{Subpopulations and ages. Upper panel: the probability
  distribution for the age of the three distinct chemo-dynamical
  subpopulations in Carina (metal rich in red, intermediate in purple,
  metal poor in blue, total in black). Lower panel: the
  age-metallicity diagram for the RGB stars in the spectroscopic
  sample; error bars are shown only for stars with a precise age
  estimate ($\delta_{\rm age}\leq1.5$ Gyr); colored points identify
  high-probability members {  (probability of being a member larger than 75 percent)} for each chemo-dynamical subpopulation.}
\label{pops_ages}
\end{figure}

In this section, we investigate how the chemo-dynamical
population splitting we have just obtained projects into the 
space of stellar ages. Grey points in Fig.~\ref{pops_ages} illustrate the 
age-metallicity relation for the RGB stars we have derived 
and presented in Sect.~\ref{sec:ages} (error-bars are also 
shown for a selection of precise measurements, where 
uncertainties are lower than $1.5\Gyr$). As previously noted, 
despite Carina's narrow RGB and the well known age-metallicity 
degeneracy which makes the uncertainty on the age of any 
single RG substantial, we identify a well defined age-metallicity 
relation on the whole population, thanks to the precise photometry 
of \citet{Stetson11} and metallicity estimates of our analysis. 

High-probability members of each subpopulation are highlighted 
in different colours in the lower panel of Fig.~\ref{pops_ages}. They do not clearly 
separate out in age, but the presence of a gradient in the mean 
ages of the three populations is clear. The upper panel of Fig.~\ref{pops_ages}
directly projects the population split into the space of stellar ages, 
by showing the distributions of the ages of members of each stellar 
subpopulation (as for plots in Fig.~\ref{pops}, distributions in Fig.~\ref{pops_ages} are 
convolved with the uncertainties of each single age measurement, which
contribute to broaden each probability distribution substantially). 
We find that the MP stellar population almost exclusively contains stars 
that are associated with the oldest isochrone in our library ($13.7\Gyr$), 
the IM has a considerable component of intermediate age stars 
($6 - 10\Gyr$), while, to continue the gradient, the MR population 
extends up to recent times.

Even though the task of individual age estimation remains challenging, 
we find that the chemodynamical division into subpopulations of the 
Carina dSph is globally compatible with the independent picture presented 
by its SF history. As the photometrically derived SF history suggests, 
Carina has experienced three major SF episodes. Indeed, we find 
independent and corroboratory evidence for this in the spectroscopy: 
three stellar subpopulations are identified, with ages compatible with 
those indicated by the SF history.

\section{Discussion and Conclusions}
\label{sec:conclusions}

We have shown that, similar to Fornax and Sculptor, the Carina dSph is
characterised by the presence of multiple stellar populations, with
distinct chemical properties, spatial distributions and kinematical
states. However, Carina appears substantially
more mixed than either Sculptor or Fornax, which is reflected in the
comparatively smaller differences among the characteristic scale
lengths (as well as velocity dispersions) of its subpopulations. The
MP population is more extended (and kinematically hotter) than the two
younger populations (respectively IM and MR). What is more notable is
that we find that the youngest MR population is more extended and at
least as kinematically hot as the IM population, with evidence for an
{\it inversion} of the usual ordering. Accordingly, the IM metallicity
population is also more concentrated than the RC population.

While metal richer stellar subpopulations are
generally more spatially concentrated and accordingly kinematically
colder, the evolutionary history of Carina partially broke this common
parallel ordering. This opens the question as to whether Carina's stellar
populations were formed with the properties we observe today -- and
then any difference from Fornax or Sculptor has to be found in the
intrinsic properties and triggers of the SF episodes -- or whether Carina 
initially had a more pronounced chemical gradient but this was subsequently 
perturbed by environmental factors.

\citet{Sales10} have shown that strong enough tidal disturbances may
be capable of homogenising the properties of multiple
subpopulations. They suggest that, by removing a substantial fraction
of the outer dark matter envelope, stripping causes the velocity
dispersion of extended populations to decrease, weakening any
kinematical difference with more concentrated tracers. Indeed,
Carina's proper motion suggests a pericenter of only a few tens of
\kpc, which is considerably smaller than either Fornax or
Sculptor. Also, the presence of an extended extratidal component 
would suggest tidal disturbance as a feasible mechanism to mix the
originally more segregated subpopulations of Carina after infall.

However, it remains unclear whether tides alone can be responsible for
Carina's present configuration. In particular, tidal effects are
strongest on more extended populations which implies that they should
naturally preserve any original outside-in ordering in the
characteristic scales of stellar subpopulations: it seems somewhat
puzzling that tides are capable of causing an inversion in the
ordering of the subpopulations as observed here.  As a consequence, it
is tempting to discuss other possible mechanism that may cause such
phenomenology.

If the current distribution of the MR population has not been altered
by tides, then the gas from which it is was originated was initially
more spread out than the gas from which the IM population was born. In
the following we list mechanisms that may be held responsible for
this.

\begin{itemize}
\item{Ram pressure: if the MR is formed after infall, gas may have been 
disturbed by the interaction with the corona of the Milky Way, 
in the form of ram pressure, perhaps also triggering star formation {  \citep[see also][on how Milky Way feedback can affect the evolution of dSph satellites, even at $100\kpc$]{Nayakshin13}}.}
\item{Stellar feedback from the IM population: if not energetic enough 
to entirely remove the remaining gas, feedback would naturally result 
in the gas being distributed on more energetic orbits {  \citep[an illustration of this effect has been recently provided by][]{El-Badry15}}. It is unclear however if the cooling necessary 
to restart star formation would newly concentrate it in the central regions.}

\item{
Interaction with an another dwarf galaxy or dark halo, {  likely before infall,} that may have triggered the star formation and perturbed the gas by dynamical interaction.}
\end{itemize}

\citet{Donghia08} have suggested low mass galaxies only light up with
star formation when belonging to groups at intermediate redshifts.
Indeed, star formation is quite challenging to achieve in such small
haloes \citep[e.g.,][]{Read06} and the larger virial mass of a group
may help dwarfs to more easily retain their gas after episodes of
intense stellar feedback \citep[e.g.,][]{Penarrubia12,Amorisco14}.
Here we note that interactions between low mass dwarfs may also
contribute in triggering and facilitating star formation, in a
systematic way over the population of low mass galaxies. Such interactions
are not unfrequent in a $\Lambda$CDM universe \citep[e.g.,][]{Deason14,Wetzel15}, and would
justify the fact that a fraction of the low mass members 
of the Local Group show signs suggesting a violent and active past \citep{Kleyna03,
Coleman04, Amorisco14b, deBoer14b}.

\section*{Acknowledgments}
We thank the anonymous referee for comments that helped improving the quality of the paper. 
It is a great pleasure to also thank Mike Irwin, Thomas de Boer and Else Starkenburg for many
discussions and precious insight. The Dark Cosmology centre is funded
by the DNRF.


\bibliographystyle{mnras} \bibliography{../Carina}


\bsp	
\label{lastpage}
\end{document}